\documentclass[superscriptaddress,floatfix,amsmath,amssymb, aps, pra,twocolumn]{revtex4-2}
  
\usepackage{graphicx}
\usepackage{dcolumn}
\usepackage{bm}
\usepackage[colorlinks]{hyperref}
\usepackage{amsthm}
\usepackage[dvipsnames]{xcolor}
\usepackage{float}
\usepackage{hyphenat}

\newcommand{\be}{\begin{equation}}
\newcommand{\ee}{\end{equation}}
\newcommand{\beq}{\begin{eqnarray}}
\newcommand{\eeq}{\end{eqnarray}}

\newcommand{\ket}[1]{\mbox{$ | #1 \rangle $}}

\newcommand{\orcid}[1]{\href{https://orcid.org/#1}{\includegraphics[width=7pt]{orcid.png}}}
  
\begin{document}

\title{Simulation of Dissipative Dynamics Without Interferometers}

\author{Fabrício Lustosa}
    \email[Correspondence email address: ]{fabricio.lustosa@ufabc.edu.br }
     \affiliation{Centro de Ciências Naturais e Humanas, Universidade Federal do ABC-UFABC, Santo André 09210-580, Brazil}

     \author{Roberto M. Serra}
    \affiliation{Centro de Ciências Naturais e Humanas, Universidade Federal do ABC-UFABC, Santo André 09210-580, Brazil}   

    \author{Luciano S. Cruz}
    \affiliation{Centro de Ciências Naturais e Humanas, Universidade Federal do ABC-UFABC, Santo André 09210-580, Brazil}
     
    \author{Breno Marques}
    \email[Correspondence email address: ]{breno.marques@ufabc.edu.br}
    \affiliation{Centro de Ciências Naturais e Humanas, Universidade Federal do ABC-UFABC, Santo André 09210-580, Brazil}

\begin{abstract}
The development of techniques that reduce experimental complexity and minimize errors is an utmost importance for modeling quantum channels. In general, quantum simulators are focused on universal algorithms, whose practical implementation requires extra qubits necessary to control the quantum operations. In contrast, our technique is based on finding a way to optimally sum Kraus operators. These operators provide us with an experimentally simplified setup where only a degree of freedom is needed to implement any one-qubit quantum channel. Therefore, using entanglement-polarized photon pairs and post-processing techniques, we experimentally built the Kraus maps, carrying out unitary and projection operations. 
\end{abstract}

\maketitle

\section{Introduction}
\label{int}
One interesting application of quantum computation is the possibility to simulate complex dynamics in a controlled quantum system \cite{georgescu2014quantum,lu2017experimental}. Over the years, many experimental implementations show the possibility of being used as simulators with different physical systems, such as NMR\cite{baugh2005experimental,xin2017quantum}, ions trap \cite{barreiro2011open,islam2011onset,schindler2013quantum}, cold atoms \cite{diehl2008quantum},  superconductors \cite{zanetti2023simulating,wang2023simulating} and, or focus in this work, single-photon \cite{almeida2007environment,salles2008experimental,jimenez2009determining,fisher2012optimal, farias2012experimental, marques2015experimental,mccutcheon2018experimental,han2021experimental,cardoso2021simulation,rojas2023non, zhan2023deterministic}
In all this experiments used an controllable quantum system to implement a complex dynamic, allowing us to understand the physical meaning underneath the dynamics, such as entanglement dynamics \cite{almeida2007environment,salles2008experimental,farias2012experimental}, energy exchange between system-environment \cite{baugh2005experimental,fisher2012optimal,cardoso2021simulation}, non-markovianity \cite{liu2011experimental,rojas2023non}, phase-transitions \cite{prosen2011nonequilibrium,islam2011onset}, quantum states engineering \cite{verstraete2009quantum}, and thermalization \cite{vznidarivc2010thermalization, vznidarivc2015relaxation} 


When utilizing a quantum optical approach, depending on the dynamics to be simulated, the setup can be difficult to build, with Sagnac (or Mach–Zehnder) interferometers
\cite{almeida2007environment,lu2017experimental} or multi-path interferometers \cite{rojas2023non}. This is needed because the system of interest is encoded
in a degree of freedom and, by using an interferometer, other photon degrees can be used as ancilla to implement the dynamics. However, one can implement dynamics without an interferometer by partitioning the acquisition time in different intervals, in which different operations are implemented \cite{marques2015experimental}. This can be seen as a decomposition of a quantum map $\epsilon$ in $N$ operators $\{M_i\}$ \cite{kraus1971general,kraus1983states}:
\begin{equation}
    \epsilon(\rho)=\sum_{i=1}^N p_iM_i\rho M_i^\dagger,
    \label{kraus}
\end{equation}
where $\rho$ is the initial state and $p_i$  is the weight of the operator $M_i$ in the dynamic with $\sum_ip_i M_i^\dagger M_i =\mathbb{I}$ , with $\mathbb{I}$ being the identity. In previous experiments~\cite{marques2015experimental} only Kraus operations which are unitary operations were used to implement the maps, limiting possible maps to be simulated. The use of partition of the detection time and ancillas allowed to simulate any dynamics of qudits (d-level quantum system)~\cite{cardoso2021simulation}. 

In this work, we present an experiment that allows the implementation of any single-qubit map, without the use of ancillary systems, inspired in the theoretical proposal in \cite{wang2013solovay}. In addition, we include projection operators in the Kraus decomposition, making possible the simulation of an arbitrary channel.

In section~\ref{sec2},  we revisited the quantum map theory, in which we based our method and defined the map decomposition to implement the dissipation dynamics. The procedure of operations needed to implement the maps are shown in section~\ref{proposal} and the experimental setup is discussed in section~\ref{sec4}. In section~\ref{results},  we present our results and, finally, in section~\ref{con} we conclude and discuss some perspectives of this work. 
\section{Open Quantum Systems}
\label{sec2}
Throughout a physical process, a quantum system under investigation 
will invariably face adverse situations influenced by the environment. 
The system-environment coupling induces decoherence, and it has 
been extensively investigated in communication protocols in the 
emerging quantum information technology domain 
\cite{caves1994quantum, holevo2012quantum, 
cariolaro2015quantum, djordjevic2021quantum, breno2023}. 

Mathematically, Completely Positive and Trace-Preserving (CPTP)
maps determine the dynamic of the system environment in quantum
channels. When the CPTP condition is satisfied, the Stinespring 
representation \cite{stinespring1955positive} incorporates the 
quantum mechanics postulates. One possibility to treat these 
dynamics is to enlarge the Hilbert space to include the environment for
the composite system evolves through unitary dynamics, according to 
$U(\rho \otimes \rho_{\text{env}})U^{\dagger}$ 
\cite{caruso2014quantum,nielsen2010quantum,petruccionebook}.
Thus, after the evolution of the composite system, one obtains the 
principal system by tracing out the environment degrees of freedom, 
taking into account $\rho^{\prime}:= \text{tr}_{\text{env}}[U(\rho 
\otimes \rho_{\text{env}})U^{\dagger}]$, where $\rho$ and 
$\rho_{\text{env}}$ represent the initial states of the system and the 
environment, respectively. The CPTP transformations allow us to 
represent the principal system dynamics as a sum of Kraus operators:
\begin{equation}
\varepsilon(\rho)=\sum_{i=1}^d K_i\rho K_i^\dagger,
\label{krausmap}
\end{equation}
where $\{K_i\}$ are Kraus operators that lie on Hilbert space $\mathcal{H}_s$ of the principal system and satisfy the completeness relation $\sum_i K_i^{\dagger} K_i = \mathbb{I}$.
In Kraus formalism, a quantum channel is characterized by a linear map $\varepsilon: \rho \rightarrow \rho^{\prime}$,  which connects density operators to density operators, obeying a maximum number of, at most, $d^2$ independent Kraus operators, with $d$ being the dimension of $\mathcal{H}_s$ \cite{nielsen2010quantum, wood2011tensor,petruccionebook}. Furthermore, as those are trace-preserving maps, the dynamics of the system under analysis are not unique and there are countless equivalent fashions to sum the Kraus operators for a single quantum channel. A straightforward way to visualize this is through the mathematical property at $\text{tr}_{\text{env}}$ when discarding the environment information it is always possible to recover:

\begin{equation}
    \sum_\mu K_\mu^{\dagger} K_\mu = \mathbb{I},
    \label{sumKraus}
\end{equation}
such that
\begin{equation}
    K_\mu = \langle \mu |U| e \rangle
    \label{constructKraus}
\end{equation}
with $\{\ket{e}\}$ being an arbitrary basis in the Hilbert space of the environment \cite{bengtsson2017geometry}. 

The trace-preserving condition brings itself an additional aspect, and indeed subtle, permitting distinct physical processes to reproduce the same output state \cite{wood2011tensor}. It has played an important role since new contributions to compose the scope of quantum mechanics theory \cite{kraus1971general, kraus1983states} up to the present in quantum information sciences \cite{holevo2012quantum}, especially work on simulations of open-system dynamics on several platforms \cite{marques2015experimental, sweke2014simulation, piani2011linear, islam2011onset, schindler2013quantum, lu2017experimental, xin2017quantum, shaham2011realizing, wei2018efficient, diehl2008quantum, wang2011quantum, georgescu2014quantum}, including quantum computers \cite{kok2007linear, verstraete2009quantum}. Therefore, this has been our base for implementing optimal Kraus operators capable of replicating any single-qubit quantum channel using photons, without the requirement for an auxiliary system.

\subsection{Single-Qubit Quantum Channel}
\label{int}\label{sec3}

\begin{figure}{h}
\includegraphics[width=0.9\columnwidth]{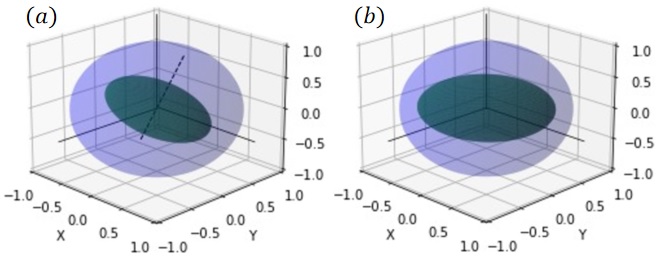}
\caption{Geometric representation of the affine map effect on the states lying on the Bloch sphere, under the assumption $\vec{\tau}=0$: (a) equation (\ref{affinemap1}) a contraction of the Bloch sphere to an ellipsoid followed by a rotation corresponding to $\vec{r} \longmapsto T \vec{r}$, (b) equation (\ref{affinemap2}) a contraction of the Bloch sphere concerning canonical axes according to $\vec{r} \longmapsto \eta \vec{r}$.}
\label{Blochsphere}
\end{figure}

We begin our demonstration by examining the effect of a quantum channel on a single qubit. A CPTP transformation can be characterized using the Bloch sphere parameterization \cite{nielsen2010quantum}. In this representation, an arbitrary density matrix can be written as
\begin{equation}
    \rho = \frac{1}{2}(\mathbb I + \vec{r}\cdot \vec{\sigma}).
    \label{qubit}
\end{equation}
The above equation is the parameterized form that associates any one-qubit state to be uniquely identified with a point on the Bloch sphere, corresponding for Bloch vector $\vec{r} = (r_x, r_y, r_z) \in \mathbb R^{3}$, where $\vec{\sigma} = (\sigma_x, \sigma_y, \sigma_z)$ denotes the column vector of Pauli matrices. Note that, the acting of a linear map $\varepsilon: \rho \rightarrow \rho^{\prime}$ will  compress the sphere into an affine transformation \cite{bengtsson2017geometry} 
\begin{equation}
    \vec{r} \longmapsto \vec{R} = T \vec{r} + \vec{\tau},
    \label{affinemap1}
\end{equation}
where $T$ is a real matrix of size $3\times3$ and $\vec{\tau}$ is a real column vector. A qubit channel, in this context, is a linear map in $\mathbb R^{3}$ that maps the Bloch sphere to a specific ellipsoid inside it. Geometrically, $T$ is a distortion matrix that modifies the shape of the sphere, and $\vec{\tau}$ is a shift vector that moves the sphere's origin \cite{wang2013solovay, bengtsson2017geometry}. Furthermore, because $T$ is a real matrix,  we always can express $T \rightarrow \mathcal{O}_1 \eta \mathcal{O}_2^{t} $ in diagonal form $\eta$  through two orthogonal transformations $\mathcal{O}_1, \mathcal{O}_2^{t} $ by singular value decomposition \cite{king2001minimal}. This allows us to express the affine map in its canonical form:  
\begin{equation}
    \vec{r} \longmapsto \vec{R} = \eta \vec{r} + \vec{\tau},
    \label{affinemap2}
\end{equation}
where the ellipsoid's shape is determined by the diagonal elements of matrix $\eta$. In this framework, pure states ($|\vec{r}|^2 = 1$) under action the map shown in equation~(\ref{affinemap2}) get mapped to the shifted ellipsoid's surface
\begin{equation}
     \Big(\frac{r_x - \tau_1}{\eta_x} \Big)^2 + \Big(\frac{r_y - \tau_y}{\eta_y}\Big)^2 + \Big(\frac{r_z - \tau_z}{\eta_z} \Big)^2 = 1,
     \label{ellipsoid}
\end{equation}
where $(\eta_x, \eta_y, \eta_x)$, and $(\tau_x, \tau_y, \tau_z)$ are the diagonal matrix elements $\eta$ and the vector components $\vec{\tau}$ respectively. The output ellipsoid based on equations (\ref{affinemap1}) and (\ref{affinemap2}) is shown in Figure~\ref{Blochsphere}. 
The transformations $\mathcal{O}_1$ and $\mathcal{O}_2^{t}$ are $3\times3$ matrices that belong to the SO(3) group and represent the ellipsoid rotations within the Bloch sphere \cite{bengtsson2017geometry}. Otherwise, the ellipsoid axes are parallel to the canonical orientation. 

The canonical version of the affine map (\ref{affinemap2}) can be used to visualize CPTP maps, whenever certain consistency requirements are met  \cite{king2001minimal, fujiwara1999one}. For example, the inequalities expressed by 
\begin{equation}
    (1 \pm \eta_z)^2 \geq (\eta_x \pm \eta_y)^2
\end{equation}
restrict the possible values of the eigenvalues $\eta_k$ for unital maps (i.e. a map 
 which preserves the identity element, $\varepsilon(\mathbb{I})=\mathbb{I}$), meaning that $\vec{\tau} = 0$ and the ellipsoid and Bloch sphere origin share the same location. These constraints are referred to as the Fujiwara-Algoet conditions \cite{fujiwara1999one}. They are necessary and sufficient conditions that determine whether the images of a CPTP map can correspond to ellipsoids within the Bloch sphere. 

On the other hand, such requirements might be extended to no-unital maps. In this case, the Algoet-Fujiwara condition becomes
\begin{equation}
      (\eta_x \pm \eta_y)^2   \leq  (1 \pm \eta_z)^2 - \tau_{z}^2,
\end{equation}
where $\eta_z = \eta_x \eta_y$, $\tau_{z}^2 = (1-\eta_x^{2})(1-\eta_y^{2})$ with $ \tau_{x} =  \tau_{y} = 0$. This allows us to represent the affine map using the trigonometric parameterization. In particular, the distortion matrix and shift vector are written in the form
\begin{equation}
\eta = \text{diag}(\cos\theta, \cos\phi, \cos\theta\cos\phi ),
\end{equation}
\begin{equation}
\vec{\tau} = (0, 0, \sin\theta \sin\phi),
\end{equation}
with $\theta \in [0, 2\pi )$, $\phi \in [0, \pi )$~\cite{ruskai2002analysis}. This parameterization can be accomplished using the Kraus decomposition, which requires only two operators
    \begin{equation}
   \mathcal{K}_1  = \left[\begin{array}{cc}
\cos\Theta &  0\\
0 &  \cos\Phi
\end{array}\right], \; \mathcal{K}_2 = \left(\begin{array}{cc}
0 &  \sin \Phi\\
\sin \Theta &  0
\end{array}\right), 
\end{equation}
with $\Theta = (\theta + \phi)/2$ and $\Phi = (\theta - \phi)/2$. An arbitrary single-qubit channel can be decomposed in terms of these operators \cite{ruskai2002analysis}. Therefore, based on the geometry of the ellipsoid, we can identify unital and non-unital maps. On the other hand, a map is considered non-unital when the ellipsoid origin does not have a fixed point $\vec{\tau} \neq 0$.  In this context, the authors in \cite{wang2013solovay} proposed an optimized quantum algorithm, which was later implemented using single photons through an interferometer setup \cite{lu2017experimental}. The algorithm simulates CPTP maps and involves a circuit with a CNOT gate and an ancilla qubit. The circuit also implements rotations $\mathcal{O}_1$ and $\mathcal{O}_2^{t}$, which are necessary to diagonalize the distortion matrix $T$. Motivated by these issues, we present a method to design distortion and shift effects carrying out rotations and projections onto photonic qubits. Following that, we will review the depolarizing (an unital map) and generalized amplitude-damping channels (a non-unital map). 
Since we show that our experimental proposal can build these two kinds of maps, then it can build a generic quantum map only changing the optical parameters.

\subsection{Depolarization Channel}
\label{IIB}
A Depolarizing channel (DP) describes the fact that the qubit remains unchanged with probability $1 - \lambda$ or changes to a maximum random state with probability $\lambda$. This can also be seen that the state to have with same probability, $\lambda / 3$, one of three error types: $\textit{Bit Flip}$, $\textit{Phase Flip}$ or $\textit{Bit-Phase Flip}$ ). One possible Kraus operator decomposition of the quantum map that describes this dynamic are
\begin{equation}
    K_0 = \sqrt{1-\lambda}\mathbb{I}, \;\;\;  K_i =\sqrt{\frac{\lambda}{3}}\sigma_i 
\end{equation}
with $(i = x,y,z)$. This channel induces the following affine map:

\begin{equation}
    \begin{pmatrix}
  x^{\prime} \\ y^{\prime} \\ z^{\prime}
\end{pmatrix} = 
\begin{pmatrix}
  1-\lambda & 0 & 0 \\
0 & 1-\lambda & 0 \\
0 & 0 & 1-\lambda 
\end{pmatrix}
\begin{pmatrix}
  x \\ y \\ z
\end{pmatrix}.
\label{DPaffine}
\end{equation}
As $\lambda \rightarrow 1$, the Bloch ball shrinks towards the origin, resulting in a completely depolarized state, which is a maximally mixed state represented by $\mathbb{I}/2$.

\subsection{Generalized Amplitude-Damping}
\label{2C}
The Generalized Amplitude-Damping channel (GAD) addresses cases in which dissipative interactions occur with the exchange of energy, including the most general example with a non-null temperature. One can use the Kraus decomposition operators:
$$
K_0 = \sqrt{\gamma}\left(\begin{array}{cc}
1 &  0\\
0 &  \sqrt{1-\lambda}
\end{array}\right), 
K_1 = \sqrt{\gamma}\left(\begin{array}{cc}
0 &  \sqrt{\lambda}\\
0 &  0
\end{array}\right) 
$$
\begin{equation}
   K_2 = \sqrt{1-\gamma}\left(\begin{array}{cc}
\sqrt{1-\lambda} &  0\\
0 &  1
\end{array}\right), K_3 =  \sqrt{1-\gamma}\left(\begin{array}{cc}
0 &  0\\
\sqrt{\lambda} &  0
\end{array}\right), 
\end{equation}
\\
where parameter $\lambda$ is the parameterized time, while $\gamma$ related to the temperature with $\lambda, \gamma$ values $\in [0,1]$ \cite{khatri2020information}. This channel has the following form in affine representation:
\begin{equation}
\begin{pmatrix}
  x^{\prime} \\ y^{\prime} \\ z^{\prime}
\end{pmatrix} = 
\begin{pmatrix}
 \sqrt{1-\lambda} & 0 & 0 \\
0 & \sqrt{1-\lambda} & 0 \\
0 & 0 & 1-\lambda 
\end{pmatrix}
\begin{pmatrix}
  x \\ y \\ z
\end{pmatrix}
+
\begin{pmatrix}
  0 \\ 0 \\ \lambda(2\gamma -1)
\end{pmatrix}
\label{GADaffine}
\end{equation}
When $\lambda \rightarrow 1$ the GAD channel drives the system to the stationary state $\rho_{\infty}= \gamma |0\rangle \langle 0| + (1-\gamma)|1\rangle \langle 1|$. It is worth noticing that when $\gamma \rightarrow 0$ or $1$, it returns to the ordinary amplitude-damping channel \cite{fujiwara2004estimation}. As $\lambda \rightarrow 1$, every point on the ellipsoid moves toward the north ($\gamma = 1$) pole or south pole ($\gamma = 0$).






\section{Proposal of Quantum Maps Without Interferometers}
\label{proposal}

In the next two subsection we will disccuss individually how we decompose the maps discussed in subsections \ref{IIB} and \ref{2C}.

\subsection{Depolarization Decomposition}

To implement single-qubit Pauli quantum channels, we consider a decomposition of working Kraus operators dependent on parameters that allow us to control the trigger probability during the operation time. The first Kraus map of interest is composed by
$$
\mathcal{M}_0 = \sqrt{p_0}\,\mathbb{I}, \;\;\;\;\mathcal{M}_x = \sqrt{p_x}\sigma_x, 
$$
\begin{equation}
    \; \mathcal{M}_y = \sqrt{p_y}\sigma_y,\, \;\; \mathcal{M}_z = \sqrt{p_z}\sigma_z,
    \label{DPdecomposition}
\end{equation}
where $p_0 + p_x + p_y + p_z = 1$ guarantees the property $\sum_{\mu}\mathcal{M}_{\mu}^{\dagger}\mathcal{M}_{\mu}=\mathbb{I}$. We will assume a pure initial state: $|\psi \rangle = \alpha |0 \rangle + \beta|1 \rangle$, so that:
\begin{equation} 
    \rho = |\alpha|^2 |0\rangle \langle 0| + \alpha \beta^* |0\rangle \langle 1| + \alpha^* \beta |1\rangle \langle 0| + |\beta|^2 |1\rangle \langle 1|.
\end{equation}
The system state evolves as 
\begin{equation}
\varepsilon(\rho)=\sum_{i=0}^z \mathcal{M}_i\rho \mathcal{M}_i^\dagger,
\end{equation}
so that the resulting quantum operation gives us
\begin{equation}
\varepsilon(\rho) = \left(\begin{array}{cc}
 \scriptstyle(p_0 + p_y)|\alpha|^2 + (p_x + p_z)|\beta|^2 &  \scriptstyle (p_0 - p_y)\alpha \beta^* + (p_x - p_z)\alpha^* \beta \\
\scriptstyle (p_y - p_z)\alpha \beta^* + (p_0 - p_y)\alpha^* \beta & 
\scriptstyle (p_y + p_z)|\alpha|^2 + (p_0 + p_y)|\beta|^2
\end{array}\right). 
\end{equation}
The three errors (bit flip, phase flip, and phase bit flip) that occur with equal likelihood are introduced from the transformations $p_0 \rightarrow 1-\lambda; \; p_x=p_y=p_z \rightarrow \lambda/3$, and we recover the equation DP channel:
\begin{equation}
\label{depolarizationEQ}
    \varepsilon(\rho) = (1-\lambda)\rho + \frac{\lambda}{3}(\sigma_x \rho \sigma_x + \sigma_y \rho \sigma_y + \sigma_z \rho \sigma_z).
\end{equation}

Noise is introduced when $\lambda >0$ and an ensemble of error-coded states represent part of the system.  In the most degrading scenario, $\lambda =3/4$, the qubit evolves such as $\rho \longmapsto \mathbb{I}/2$ \cite{djordjevic2021quantum, nielsen2010quantum}. Notice that each of the four operations in equation (\ref{DPdecomposition}) can be implemented using combinations of two half-wave plates when the qubit is encoded in the polarization (more details in section~\ref{sec4}).


\subsection{Generalized Amplitude-Damping Decomposition}
\label{2B}
In this next example, the optimal Kraus operators are 
$$
 \mathcal{M}_0 = \sqrt{p_0}\,\mathbb{I}, \;\;\;\;\;\;  \mathcal{M}_1 = \sqrt{p_1}|0\rangle \langle 0|, \;\;\;\;\;\;\;\; \mathcal{M}_2 = \sqrt{p_2}|1\rangle \langle 1|
$$
\begin{equation}
     \mathcal{M}_{3} = \sqrt{p_3}|0\rangle \langle 1|, \;\;\;\; \mathcal{M}_{4} = \sqrt{p_4}|1\rangle \langle 0|,
     \label{GADdecomposition}
\end{equation}
with $p_0 + p_1 + p_2 + p_3 + p_4 = 1$. Applying the map described by the Kraus decomposition (\ref{GADdecomposition}), we get
\begin{equation}
\varepsilon(\rho) =
\scriptscriptstyle\left(\begin{array}{cc}
|\alpha|^2(p_0 + p_1) + |\beta|^2p_3  &  \alpha \beta^* p_0\\
\alpha^* \beta p_0 &  |\beta|^2(p_0+p_2) + |\alpha|^2p_4
\end{array}\right).
\end{equation}
In this other example, $p_0$, $p_1$, $p_2$, $p_3$, and $p_4$ fit as: $p_0 \rightarrow \sqrt{1-\lambda}, \; p_1 \rightarrow  1 - \lambda +\lambda \gamma + \sqrt{1-\lambda}, \; p_2 \rightarrow  1- \lambda \gamma  -  \sqrt{1-\lambda}, \; p_3 \rightarrow \lambda \gamma\;\; \text{and} \;\; p_4 \rightarrow \lambda -  \lambda \gamma$. We get
\begin{equation}
\label{dampingEQ}
\varepsilon(\rho) =
\left(\begin{array}{cc}
\scriptstyle |\alpha|^2(1-\lambda + \lambda\gamma) + |\beta|^2\lambda \gamma & \scriptstyle   \alpha \beta^* \sqrt{1-\lambda}\\ \scriptstyle
\alpha^* \beta \sqrt{1-\lambda} & \scriptstyle |\alpha|^2(1 - \gamma)\lambda + |\beta|^2(1- \lambda \gamma)
\end{array}\right).
\scriptscriptstyle
\end{equation}
 If we set $p_4 = 0$, the Kraus map is reduced to the ordinary \textit{amplitude-damping channel} composed by operators: $\{ \mathcal{M}_0, \mathcal{M}_1, \mathcal{M}_2, \mathcal{M}_{3} \}$. Following the same reasoning, $p_3 = 0$ removes $\mathcal{M}_{3}$, and the Kraus map mimics the \textit{dephasing channel}.  Note that, the operators $\mathcal{M}_{1}$ and $\mathcal{M}_{2}$ shown in this proposal can be implemented using polarizer  or a combination of half-wave plate and polarization. In the case of $\mathcal{M}_{3}$ and $\mathcal{M}_{4}$ we will give more details in the next section.




\section{Experimental Setup}
\label{int}\label{sec4}

\begin{figure}
\includegraphics[width=0.69\columnwidth]{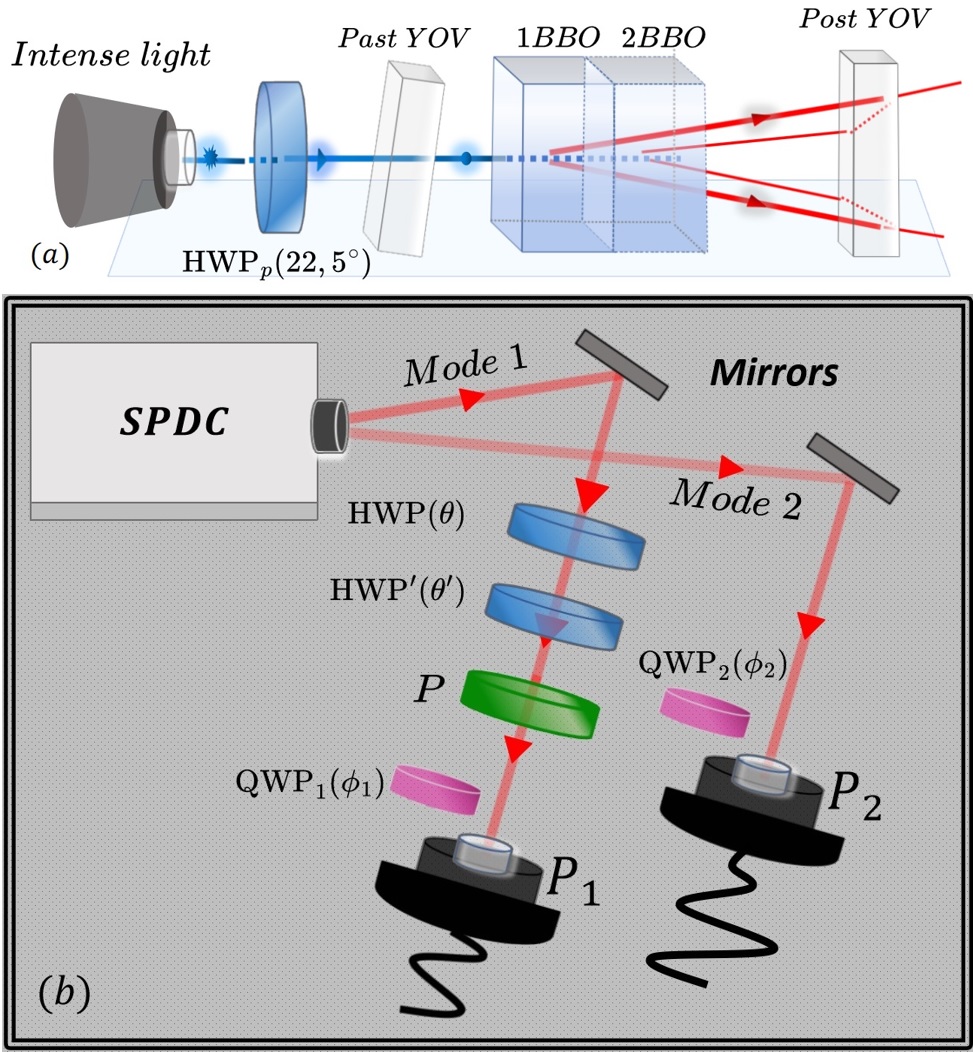}
\caption{Experimental setup to Implement quantum maps: (a) Infrared photon pairs are generated by a pump at diagonal polarization after $\text{HWP}_p$ set at $22,5^\circ$ so that down-conversion occurs either crystal first or second. (b) Two HWP$^{(\prime)}$ plates and a polarizer $P$ perform unitary operations and projections in mode 1 to generate the Kraus operations needed to implement the maps. Next, the polarizers $P_{1(2)}$ coupled with mono-mode fibers and two quarter-wave plates QWP$_{1(2)}$ are used to carry out the apparatus to  make the state tomography.}
\label{source}
\end{figure}

The initial system state is prepared as close as possible to a maximally entangled Bell state in polarization through an spontaneous parametric down-conversion (SPDC) process.  For our experiments, we used an experimental setup based on the SPDC source designed by Qutools company \cite{qutoolsquedmanual}. The device is designed in two basic setups, as illustrated in Figure~\ref{source}, where the first part is composed of two nonlinear crystals, a laser beam, and walk-off components required to complete the preparation of the entangled state through SPDC process, Figure~\ref{source}(a). In the next one, Figure~\ref{source}(b), the mirrors direct the down-converted photons into a single spatial mode defined by an optical fiber. Then, the photons are detected in coincidence by single-photon detectors.

Initially, a beam from a diode laser, operating at a wavelength of $405~nm$, pumps a $\beta$-barium borate (BBO) crystal, which has been cut out adequately for carrying out collinear type-I phase matching, producing photon pairs with wavelengths of $\lambda_1=\lambda_2=810~nm$. Due to that phase-matching condition, such photons are created with the same polarization and are orthogonal to the higher-energy ultraviolet photon, which makes it possible to determine the photon polarization of lower energy. 

The previously mentioned feature is largely exploited to get polarization entanglement. The common technique involves placing two identical BBO crystals in an abreast way so that their optical axes are held orthogonal to each other. For instance, A half-wave plate (HWP) set at $22,5^{\circ}$, fixed before the nonlinear crystals, turns the pump photon polarization $|H \rangle, |V \rangle$  into $1/\sqrt{2}( |H \rangle \pm  |V \rangle )$, such that the SPDC process takes place either on the first or on the second crystal \cite{garrison2008quantum}. Moreover, a pair of YVO crystals is used to compensate for some temporal and spatial effects that ill the system's coherence \cite{qutoolsquedmanual}. This configuration prevents any possibility of determining the crystal of origin for the down-converted photons; accordingly, our source produces two-photon entangled states close to the state $|\Phi^{-}\rangle = \frac{1}{\sqrt{2}}[|H_1 \rangle |H_2 \rangle - |V_1 \rangle |V_2 \rangle]$.


Although our system is in a two-qubit state, noise-simulation setup only runs single-qubit maps, but we use a two-qubit system to characterize the quality of our implementation. This is also illustrated in Figure \ref{source}(b), where the optical elements required for performing each operation $\mathcal{M}_{i} \rho \mathcal{M}_{i}^{\dagger}$ act solely on photons in mode 1. We use two half-wave plates, HWP and HWP$^{\prime}$, along with a polarizer $P$, to implement unitary and projection operations from decompositions (\ref{DPdecomposition}) and (\ref{GADdecomposition}). In the geometric picture, the HWP-plates execute rotations on the Bloch sphere, while $P$ projects the qubit at the north and south poles. In addition, we obtain the resulting density matrix for each operation element $\mathcal{M}_{i}$ applied using standard quantum state tomography (QST)~\cite{toninelli2019concepts}. In this way, photon polarization analysis is carried out with moving quarter-wave plates QWP$_{1(2)}$ plus polarizers $P_{1(2)}$ coupled with mono-mode fibers connected with detectors, which register the photon-pair coincidence rate. 
As discussed before, the vast majority of the work on quantum channels implementations,  in single-photon systems, has been focused on interferometer assembly \cite{salles2008experimental, piani2011linear, jeong2013experimental, liu2017experimental}. In this context, the spatial modes of an interferometer are auxiliary systems that represent the freedom degrees of the environment system. Although photonic technologies offer various ways for encoding two-level systems \cite{simon2017quantum}, accurate control for tuning the parameters of quantum hardware can make such processes extremely complex or intractable to accomplish  \cite{aspuru2012photonic}. Differently, our approach avoids the ``obligation" to represent the external physical system suggested by the Kraus-Stinespring criterion, and only the polarization degree is used throughout the process. Thus, our strategy for simulating the maps consists of experimentally building output states $p_{i}^{-1} \mathcal{M}_{i} \rho \mathcal{M}_{i}^{\dagger}$ and operationally preparing the weights $p_i$ via post-processing. The table \ref{table} shows the sequence of how the optical elements are combined to generate the outputs. As a last step of the simulation scheme, we subdivide the total time of operation into different intervals for each $ \mathcal{M}_{i} \rho \mathcal{M}_{i}^{\dagger}$ using the post-processing technique considering the ratios given shown in equations~\ref{depolarizationEQ} and~\ref{dampingEQ}.

\begin{figure}
\includegraphics[width=0.9\columnwidth]{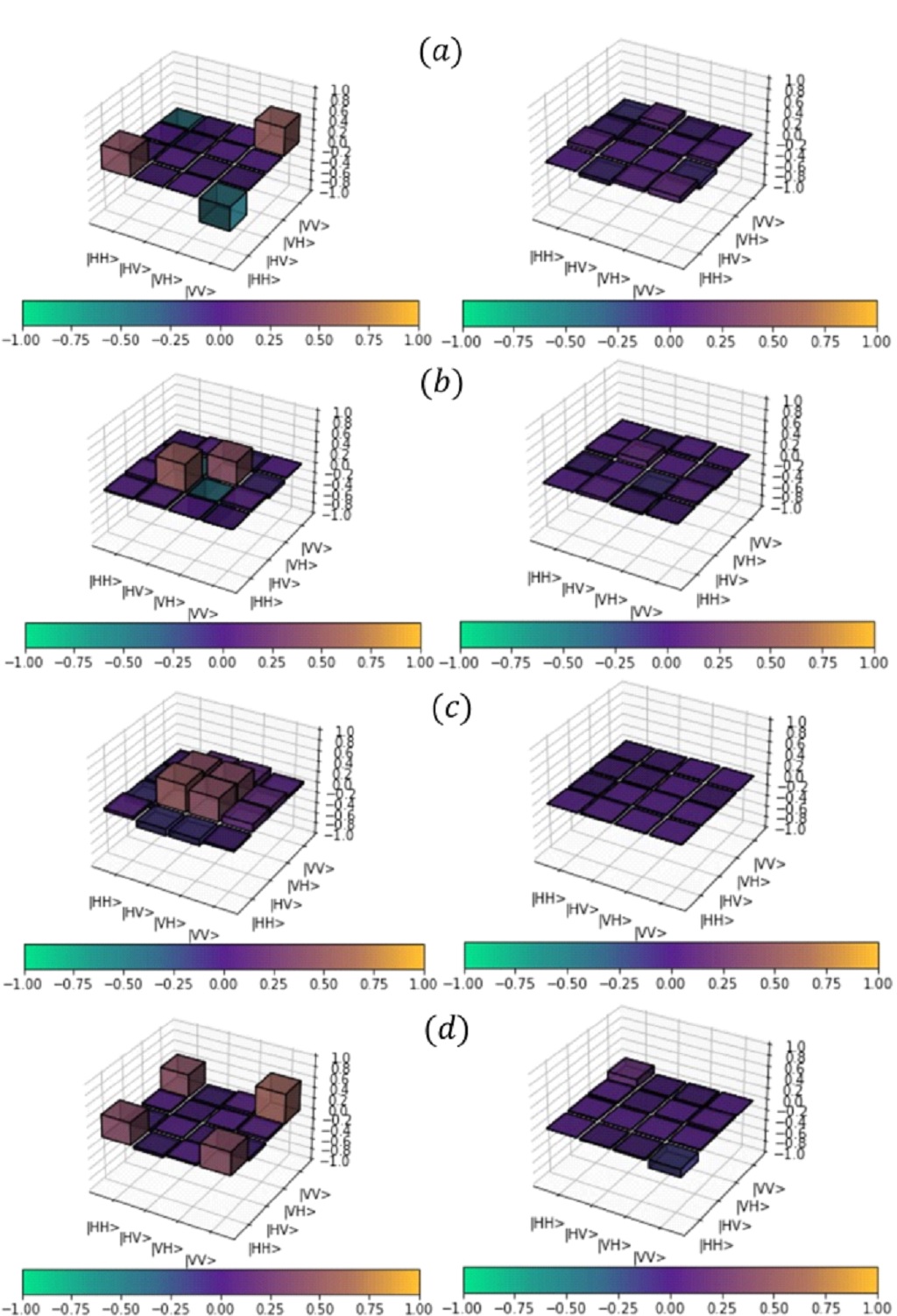}
\caption{Density matrix reconstruction via quantum state tomography (QST). (a)-(d) the real and imaginary parts of the density matrix of each operation $\mathcal{M}_i$ in (\ref{DPdecomposition}) outlined in table \ref{table}. The fidelities for the four Bell states are 0.93, 0.90, 0.89, and 0.92, respectively.}
\label{DPtomography}
\end{figure}

\begin{figure}
\includegraphics[width=0.9\columnwidth]{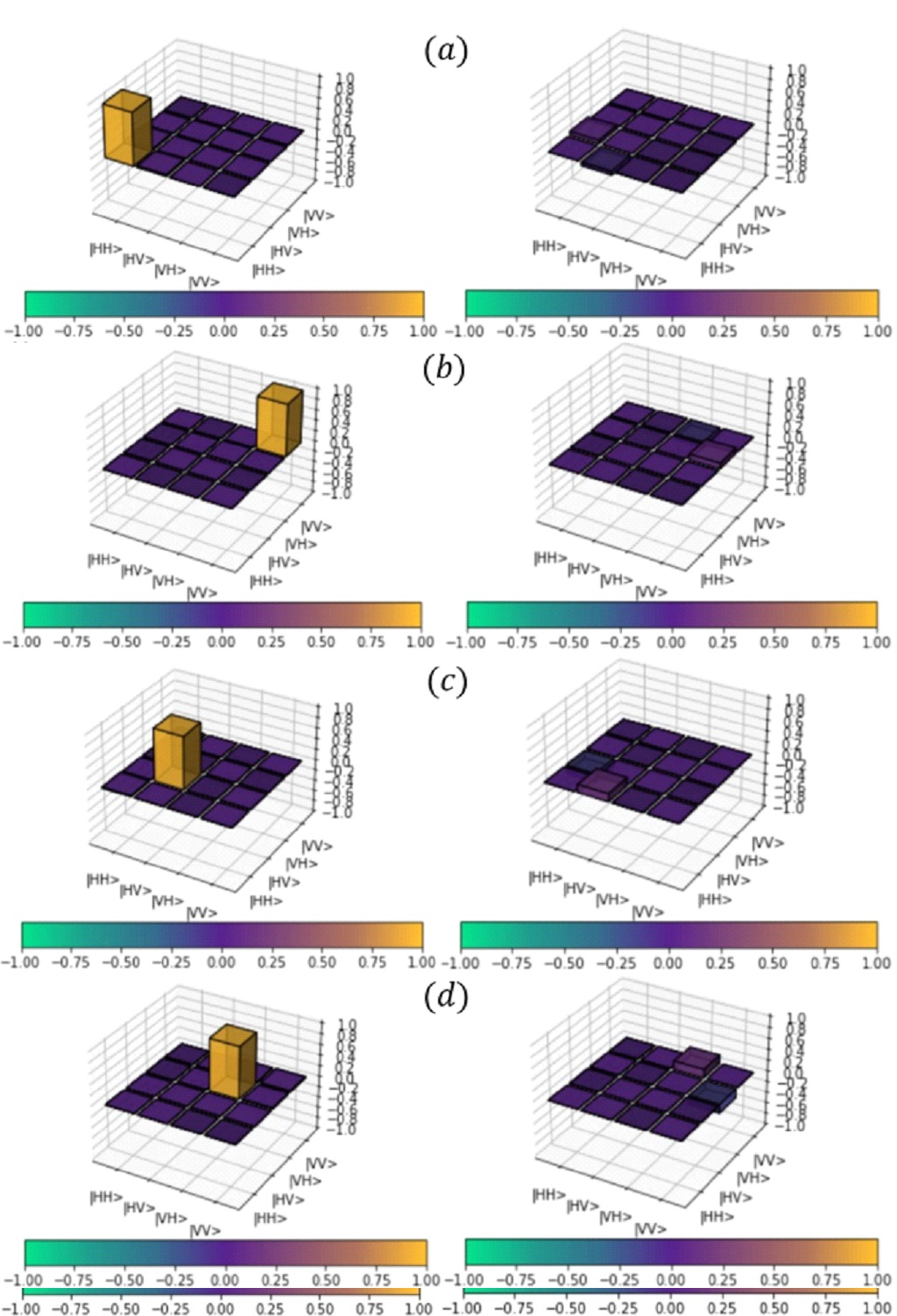}
\caption{Density matrix reconstruction via quantum state tomography (QST). (a)-(d) the real and imaginary parts of the density matrix of four operations $\mathcal{M}_i \; (\text{with} \; i = 1,2,3,4.)$ in (\ref{GADdecomposition}) outlined in table \ref{table}.  Note that operation $\mathcal{M}_0$ is common in both channels and therefore its result is just plotted onto Figure \ref{DPtomography}.}
\label{GADtomography}
\end{figure}
\begin{table}[t]
\begin{center}
\begin{tabular}{| l | r | r | r | r | r | r |}
\hline
\hline
\multicolumn{5}{|c|}{\textbf{Depolarizing Channel}}
\\
\hline
\hline
$\scriptstyle \text{Input}$ & $\hspace{0.1cm}\scriptstyle p_k^{-1}\mathcal{M}_k\hspace{0.1cm}$ & $\scriptstyle \text{implementation}$ & $\scriptstyle \text{Mode\;1}\hspace{1.0cm}$ & $\scriptstyle \text{Output}\hspace{0.25cm}$\\
\hline
& $\scriptstyle p_0^{-1}\mathcal{M}_{0}\hspace{0.1cm}$ & $\scriptstyle \mathbb{I}\otimes\mathbb{I}\hspace{0.7cm}$ & - \hspace{1.2cm}
& $\scriptstyle |\Phi^{-}\rangle \hspace{0.25cm}$
\\
$\scriptstyle |\Phi^{-}\rangle$ & $\scriptstyle p_x^{-1}\mathcal{M}_x\hspace{0.1cm}$& $\scriptstyle \sigma_x\otimes\mathbb{I}\hspace{0.7cm}$ & $\scriptstyle \text{HWP}(45^\circ) \hspace{0.9cm}$  & $\scriptstyle |\Psi^{-}\rangle \hspace{0.25cm}$
\\
& $\scriptstyle p_y^{-1}\mathcal{M}_y\hspace{0.1cm}$ & $\scriptstyle \sigma_x\sigma_z\otimes\mathbb{I}\hspace{0.7cm}$ & $\scriptstyle \text{HWP}(0^\circ) + \scriptstyle \text{HWP}^{\prime}\scriptstyle(45^\circ)\hspace{0.3cm}$ &  $ \scriptstyle |\Psi^{+}\rangle \hspace{0.25cm}$
\\
&  $\scriptstyle \scriptstyle p_z^{-1}\mathcal{M}_z\hspace{0.1cm}$ & $\scriptstyle \sigma_z \otimes \mathbb{I} \hspace{0.7cm}$  & $\scriptstyle \text{HWP}(0^\circ) \hspace{1.0cm} $ &  $\scriptstyle |\Phi^{+}\rangle \hspace{0.25cm}$
\\
\hline
\hline
\multicolumn{5}{|c|}{\textbf{Generalized Amplitude Damping Channel}}
\\
\hline
\hline
$\scriptstyle \text{Input}$ & $\hspace{0.1cm}\scriptstyle p_k^{-1}\mathcal{M}_k\hspace{0.1cm}$ & $\scriptstyle \text{implementation}$ & $\scriptstyle \text{Mode\;1}\hspace{1.0cm}$ & $\scriptstyle \text{Output}\hspace{0.25cm}$\\
\hline
& $\scriptstyle p_0^{-1}\mathcal{M}_{0}\hspace{0.1cm}$ & $\scriptstyle \mathbb{I}\otimes\mathbb{I}\hspace{0.7cm}$ & - \hspace{1.3cm} &  $\scriptstyle |\Phi^{-}\rangle \hspace{0.25cm}$
\\
 & $\scriptstyle p_1^{-1}\mathcal{M}_1\hspace{0.1cm}$& $\scriptstyle |H\rangle \langle H|\otimes\mathbb{I}\hspace{0.7cm}$ & $\scriptstyle \text{ Polarizer\;in\;H} \hspace{0.7cm}$ &  $\scriptstyle |H\rangle_1|H\rangle_2$
\\
$\scriptstyle |\Phi^{-}\rangle$ & $\scriptstyle p_2^{-1}\mathcal{M}_{2}\hspace{0.1cm}$ & $\scriptstyle |V\rangle \langle V|\otimes\mathbb{I} \hspace{0.7cm}$ & $\scriptstyle \text{ Polarizer\;in\;V} \hspace{0.7cm}$ &  $\scriptstyle |V\rangle_1|V\rangle_2$
\\
& $\scriptstyle p_3^{-1}\mathcal{M}_{3}\hspace{0.1cm}$ & $\scriptstyle |H\rangle \langle V| \otimes\mathbb{I}\hspace{0.7cm}$  & $\scriptstyle \text{HWP}(45^{\circ}) + \text{ Polarizer\;in\;H}$   & $\scriptstyle |H\rangle_1|V\rangle_2$
\\
& $\scriptstyle p_4^{-1}\mathcal{M}_{4}\hspace{0.1cm}$ & $\scriptstyle |V\rangle \langle H| \otimes\mathbb{I}\hspace{0.7cm}$  & $\scriptstyle \text{HWP}(45^{\circ}) + \text{ Polarizer\;in\;V}$   & $\scriptstyle |V\rangle_1|H\rangle_2$
\\
\hline
\end{tabular}
\centering
\caption{\label{table}The layout of the sequence of operations.Each Kraus operator present in the desired single-qubit channel is implemented by the optical elements placed along mode 1.}
\end{center}
\end{table}

\section{Results and Discussion}
\label{results}
The preceding discussion can be formalized from the idea that our physical system is an ensemble of entangled photons emitted over a certain time interval. In this way, the resulting output state $\varepsilon(\rho)$ is reproduced during the entire time detection:
\begin{equation}
 \Delta T = \sum_{i=0}^{n}\Delta t_i,
\end{equation}
where each $\Delta t_i$ can be different. In our case, the stages of dynamic can be determined by subdividing time detection $\Delta T$ into smaller periods   $\Delta t_i$ to specify the weight for which each operator $\mathcal{M}_{i}$ will act on the qubit \cite{marques2015experimental}, where $p_i=\Delta t_i/\Delta T$. In the affine representation of channel examples~(\ref{DPaffine})and~(\ref{GADaffine}) the value of $\lambda$ allows us to map the qubit during the whole dynamic choosing correctly the values of $\{\Delta t_i\}$. We note that, to the simulation of generalized amplitude-damping in the fashion described here, it is necessary to use negative values of $p_i$. In this case, instead of summing the counts, we subtract the results in that time interval, in a similar way of characterizing a mean value of an observable, and the detecting counts of a projection with negative eigenvalue are subtracted. 

In the present implementation, we do not have automated optical mechanics that allow us to apply different Kraus operators in a single run.  Instead, we made all the measurements in $\Delta T =10~s$ for each Kraus operator and measurement of the quantum tomography, afterward, we combined the Kraus operators and $\{p_i\}$ weights using post-processing to estimate the output state.

To demonstrate that our strategy can be used to simulate a given map we compared our results and the theoretical dynamics using 3 figures of merit: the experimental reconstructed state and theoretical states fidelity, entanglement, and purity. The fidelity between two states $\rho_1$ and $\rho_2$ is given by $F(\rho_1,\rho_2)=\text{tr}[\sqrt{\rho_1}\rho_2\sqrt{\rho_1}]$, the concurrence \cite{chen2012estimating} was used to quantify the entanglement and the purity is given by $P=\text{tr}[\rho^2]$.


\subsection{Depolarizing channel Implementation}


Werner states are frequently produced through the DP channel applied in a Bell state and we expect that we see a similar dynamic. Using the plates HWP$^{(\prime)}$ in our experimental setup as shown in the Table~\ref{table} we implement the following outputs:
$$
 |\Phi^{-} \rangle  \rightarrow \Bigg( \frac{\mathcal{M}_0}{\sqrt{p_{0}}} \otimes \mathbb{I}\Bigg)|\Phi^{-} \rangle, \; |\Psi^{-} \rangle  \rightarrow \Bigg( \frac{\mathcal{M}_x}{\sqrt{p_{x}}} \otimes \mathbb{I}\Bigg)|\Phi^{-} \rangle
$$
\begin{equation}
   |\Psi^{+} \rangle  \rightarrow \Bigg( \frac{\mathcal{M}_y}{\sqrt{p_{y}}} \otimes \mathbb{I} \Bigg)|\Phi^{-} \rangle, \; |\Phi^{+} \rangle  \rightarrow \Bigg( \frac{\mathcal{M}_z}{\sqrt{p_{z}}} \otimes \mathbb{I}\Bigg)|\Phi^{-} \rangle.
\end{equation}

To show if we are implementing correctly the map, we made the tomographic reconstructions of the real and imaginary components of the density matrix for the four transformations above, as illustrated in Figure \ref{DPtomography}. Applying the weights $p_i$ according to the decomposition (\ref{DPdecomposition}), we obtain
$$
\varepsilon(|\Phi^{-} \rangle \langle \Phi^{-}|) \rightarrow 
\sum_{i=0}^z \mathcal{M}_i |\Phi^{-} \rangle \langle \Phi^{-}| \mathcal{M}_i^\dagger.
$$
Taking into account the reasoning presented in Sec.\ref{proposal} that $p_0 = 1-\lambda$ and $p_x = p_y = p_z = \lambda/3$, the pure state evolves to
\begin{equation}
    |\Phi^{-} \rangle \langle \Phi^{-}| \longmapsto \Bigg(1 - \frac{4}{3}\lambda \Bigg)|\Phi^{-} \rangle \langle \Phi^{-}| + \frac{\lambda}{3}\mathbb{I \otimes I}.
\end{equation}
From the relation $\lambda = \Delta t_{i} / \Delta T$, with $\{i=x,y,z\}$,  we quantify the effective time for the operations $\mathcal{M}_i$. For instance, by assigning the acquisition time $\Delta T$, the effect of the DP channel occurs at intervals ranging from 0 to $\Delta T$:

$$
\lambda =0, \quad\vec{t}=\begin{pmatrix} \Delta T\\0\\0\\0  
\end{pmatrix}\; \rightarrow  \; |\Phi^{-}\rangle \langle \Phi^{-}|,\nonumber\\
$$
$$
\lambda =\frac{1}{4}, \quad \vec{t}=\begin{pmatrix}  0.75\\0.08\overline{3}\\0.08\overline{3}\\0.08\overline{3}  
\end{pmatrix}\; \rightarrow  0.\overline{6}|\Phi^{-}\rangle \langle \Phi^{-}|+0.08\overline{3}\; \mathbb{I \otimes I}\\
$$

$$
\lambda =\frac{1}{2}, \quad \vec{t}=\begin{pmatrix}  0.5\\0.1\overline{6}\\0.1\overline{6}\\0.1\overline{6}  
\end{pmatrix}\; \rightarrow  0.\overline{3}|\Phi^{-}\rangle \langle \Phi^{-}|+0.1\overline{6}\; \mathbb{I \otimes I},\\
$$
\begin{equation}
\lambda =\frac{3}{4}, \quad \vec{t}=\begin{pmatrix}  0.25\\0.25\\0.25\\0.25  
\end{pmatrix}\; \rightarrow  \frac{1}{4}\mathbb{I \otimes I}.    
\end{equation}
At $\Delta t_i=0$, the physical system is in a pure state, and during the acquisition time, only $\mathcal{M}_0$ acts during all acquisition time $\Delta T$. However, as the acting time of operators $\mathcal{M}_i \; (i = x,y,z)$ increases, they become a significant part of the acquisition time. Consequently, errors will be introduced gradually until that system reaches the most degrading scenario in $\Delta t_i = 3\Delta T /4$. At this point, the system will be mapped to the maximally mixed density operator. 

\begin{figure}
\centering
\includegraphics[width=1\columnwidth]{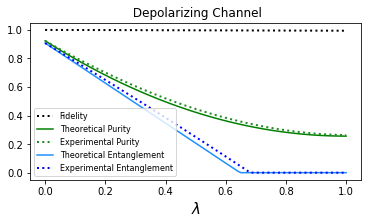}
\caption{Fidelity, purity, and entanglement theoretical (full line) and experimental (dotted line) curves for the DP.}
\label{DPlines}
\end{figure}

Figure \ref{DPlines}(a) outlines the depolarizing of a qubit, or how to build Werner states controlling the action of the Kraus operators $\mathcal{M}_i$ into temporal partitions. 
When,  $\Delta t_i \rightarrow 0 \; ( \lambda \rightarrow 0)$ the green curve denotes that the system is pure, or otherwise, the system can be into a maximally mixed state whenever $\Delta t_i \rightarrow 3\Delta T /4 \; (\lambda \rightarrow 1)$. Therefore, when a mode 1 photon is sent to the DP channel, the photon-pair entanglement has a sudden death of entanglement in our experiment at $\lambda \approx 0.65$, very close to the ideal Werner state, that is $\lambda=2/3$ and for our initial state, that has $\lambda\approx0.62$.


\subsection{Generalized Amplitude Damping Channel Implementation}
\label{5B}

In the next step, we use a plate HWP and a polarizer $P$ to obtain the output states in the scheme depicted in Figure \ref{source}(b). This results in the following transformations on the joint state of modes 1 and 2:
\begin{eqnarray}
 |\Phi^{-} \rangle  \rightarrow \Bigg( \frac{\mathcal{M}_0}{\sqrt{p_{0}}} \otimes \mathbb{I}\Bigg)|\Phi^{-} \rangle,  \nonumber\\
|H_1 \rangle | H_2
 \rangle  \rightarrow \Bigg( \frac{\mathcal{M}_1}{\sqrt{p_{1}}} \otimes \mathbb{I}\Bigg)|\Phi^{-} \rangle, \nonumber\\
  |V_1 \rangle | V_2
 \rangle  \rightarrow \Bigg( \frac{\mathcal{M}_2}{\sqrt{p_{2}}} \otimes \mathbb{I} \Bigg)|\Phi^{-} \rangle,\\
|H_1 \rangle | V_2
 \rangle  \rightarrow \Bigg( \frac{\mathcal{M}_3}{\sqrt{p_{3}}} \otimes \mathbb{I}\Bigg)|\Phi^{-} \rangle, \nonumber\\|V_1 \rangle | H_2
 \rangle  \rightarrow \Bigg( \frac{\mathcal{M}_4}{\sqrt{p_{4}}} \otimes \mathbb{I}\Bigg)|\Phi^{-} \rangle.\nonumber
\end{eqnarray}
The tomographic reconstructions of the four real and imaginary density matrix components of the five outputs necessary to implement the channel are shown in Figure \ref{GADtomography}. It should be noted that the tomography result of the source depicted in Figure \ref{DPtomography} indicates an identity operation and was used to run both channels.

\begin{figure}
\centering
\includegraphics[width=0.8\columnwidth]{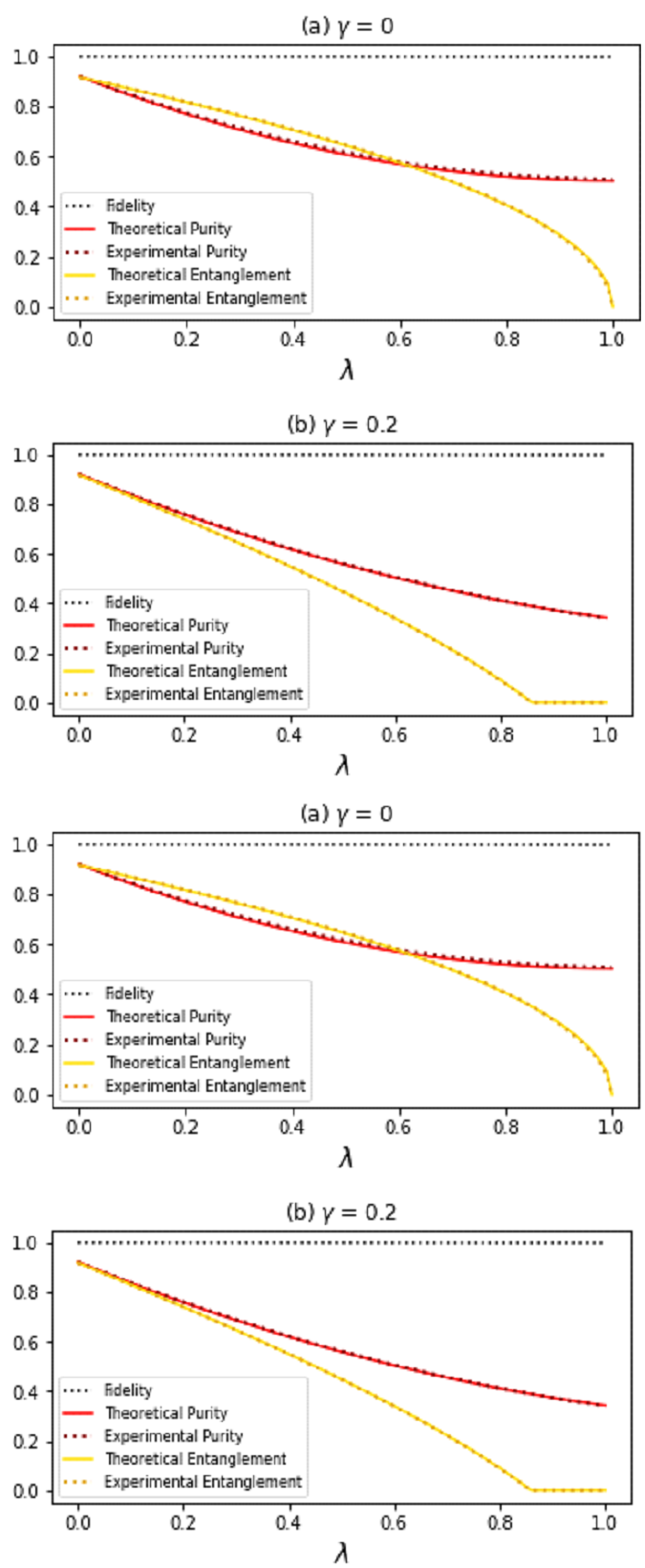}
\caption{Fidelity, purity, and entanglement theoretical (full line) and experimental (dotted line) curves for the DP and GAD channels (a) depolarizing channel and (a)-(d) generalized amplitude channel for some different temperature values: $\gamma = 0, \; 0.2 \;, \; 0.4 \; \text{and} \; 1$.}
\label{GADlines}
\end{figure}

In section \ref{2B}, we have discussed how the weights $p_i$ are adjusted to reproduce the GAD channel. Therefore, when the qubit undergoes the map action (\ref{GADdecomposition}) the Bell $|\Phi^{-} \rangle \langle \Phi^{-}|$ evolves to 
$$
 \frac{1}{2}\big[  (1-\lambda + \lambda \gamma)|H_1  H_2
 \rangle  \langle H_1  H_2|  +  \lambda \gamma|H_1  V_2
 \rangle  \langle H_1  V_2|
$$
$$
 (1 - \gamma)\lambda |V_1  H_2
 \rangle  \langle V_1  H_2| + (1 - \lambda \gamma) |V_1  V_2
 \rangle  \langle V_1  V_2| 
$$
\begin{equation}
    -\sqrt{1-\lambda}(|H_1  H_2
 \rangle  \langle V_1  V_2| + |V_1  V_2
 \rangle  \langle H_1  H_2| ) \big]
\end{equation}
or
$$
|\Phi^{-} \rangle \langle \Phi^{-}| \rightarrow \frac{1}{2}\left(\begin{array}{cccc}
1-\lambda + \lambda \gamma &  0  &  0 & -\sqrt{1-\lambda}\\
0 &  \lambda \gamma  &  0 & 0\\
0 &  0  &  (1 - \gamma)\lambda & 0\\
-\sqrt{1-\lambda} &  0  &  0 & 1 - \lambda \gamma\\
\end{array}\right).
$$
We set $\Delta T$ for the acquisition time and $p_i = \Delta t_i / \Delta T$, $i=0,1,2,3,4$. In this simulation, for a chosen temperature, defined by $\gamma$), we establish dynamic stages by $\vec{t}^T=(\Delta t_0,\Delta t_1,\Delta t_2,\Delta t_3,\Delta t_4)$ as follows:

$$
\lambda =0 \quad\vec{t}=\begin{pmatrix} \Delta T\\0\\0\\0\\0  
\end{pmatrix}\; \rightarrow  \; |\Phi^{-}\rangle \langle \Phi^{-}|,\nonumber\\
$$
$$
\scriptstyle \lambda = 0.1 \quad \scriptstyle \vec{t}=\begin{pmatrix} \scriptstyle 0.95\\ \scriptstyle 1.85 + 
 0.1\gamma\\ \scriptstyle 0.05 - 0.1\gamma\\ \scriptstyle 0.1\gamma\\ \scriptstyle (1-\gamma)0.1
\end{pmatrix}\; \rightarrow  \; \scriptstyle   \frac{1}{2}\left(\begin{array}{cccc}  
 \scriptstyle 0.9 + 0.1 \gamma &  \scriptstyle 0  &  \scriptstyle 0 & \scriptstyle -\sqrt{0.9}\\
\scriptstyle 0 &  \scriptstyle 0.1 \gamma  &  \scriptstyle 0 & \scriptstyle 0\\
\scriptstyle 0 &  \scriptstyle 0  & \scriptstyle (1 - \gamma)0.1 & \scriptstyle 0\\
\scriptstyle -\sqrt{0.9} & \scriptstyle  0  & \scriptstyle  0 & \scriptstyle 1 - 0.1 \gamma\\
\end{array}\right).
$$

$$
\vdots \;\;\;\;\;\;\;\;\;\;\;\;\;\;\; \vdots \;\;\;\;\;\;\;\;\;\;\;\;\;\;\;  \vdots
$$

\begin{equation}
\lambda = 1, \quad \vec{t}=\begin{pmatrix} 0\\\gamma\\1-\gamma\\\gamma\\1-\gamma 
\end{pmatrix}\; \rightarrow    \frac{1}{2}\left(\begin{array}{cccc}  
 \scriptstyle \gamma &  0  &  0 &  0\\
0 &  \scriptstyle \gamma  &  0 & 0\\
0 &  0  & \scriptstyle 1 - \gamma & 0\\
0 &  0  &  0 & \scriptstyle 1 - \gamma \\
\end{array}\right).
\end{equation}

or equivalently 

\begin{equation}
\begin{pmatrix}
  \gamma & 0 \\
0 & 1-\gamma  \\ 
\end{pmatrix}
\otimes
\begin{pmatrix}
  1/2 & 0 \\
0 & 1/2  \\ 
\end{pmatrix},
\label{GADoutput}
\end{equation}
The final expression means that the channel drives 
 the photon in the mode 1 to the stationary state $\rho_{\Delta t\rightarrow 10}= \gamma |H_1\rangle \langle H_1| + (1-\gamma)|V_1\rangle \langle V_1|$, when we consider the full ensemble of measurements. It can be verified by tracing out the degrees of freedom of the mode-2 photon. Figures \ref{GADlines} (a)-(d) illustrates the evolution discussed in Sec.\ref{5B} for four different temperatures: $\gamma = 0$ ($1$) implies that the channel simulates an amplitude damping case with a null temperature (infinite temperature) \cite{fujiwara2004estimation},  while for $\gamma = 0.2$ and $\gamma=0.4$, we have the termalization of the system for a finite temperature. Moreover, only in finite temperatures, we can see the sudden death of entanglement \\textcolor{blue}{[\textit{acho interessante citar uma Ref. aqui}]}.

A situation comparable to the DP channel, i.e., the system entanglement comes down to zero before achieving $ \lambda = 1$. It turns out that when $\lambda \rightarrow 1$ for fixed values of $\gamma$ around $1/2$ for both modes 1 photon and the photon pair, they will be in a state of maximum mixing. On the other hand, in the ordinary case ($\gamma = 0$ or $1$) only the mode-2 photon will be in the state of maximum mixing. The mode 1 photon will be mapped to the pure state,  but the system as a whole will be in $\mathbb{I \otimes I}/2$.

\newpage

\section{Conclusions}
\label{con}
We have shown how our apparatus efficiently implements the Kraus operators $\mathcal{M}_i$, allowing us to model CPTP maps without using interferometric schemes. We employ the post-processing technique as an instrument to analyze the experimental outputs and plot graphics.  Although we have simulated the quantum maps using post-processing, we do want to emphasize that our technique is entirely feasible to implement by adding or subtracting photon-pair coinciding count rates during the experiment in a further setup. This can be accomplished using optical components mounted on automated platforms to perform the operations in Table I during a suitable time interval in the complete experimental run. We used as examples of our method the depolarization channel as an unital map. Arbitrary unital maps can also be built by changing the weight of the Pauli matrices and rotating them, which changes the elliptical deformation axis. We also showed an example of a non-unital map in the generalized amplitude damping channel case. Depending on different parameter $\gamma$, we can have any asymptotic state in the $\sigma_z$ axis of the Bloch sphere. Applying a unitary transformation in the map operators, $\{\mathcal{M}'_{i}\}=\{U\mathcal{M}_{i}U^{\dagger}\}$, one can have a new map with the asymptotic state being any state in the Bloch sphere. Therefore, results indicate a simple experimental setup that can lead to enhanced control of quantum states and scaling it up to larger systems. This not only reduces any spatial complexity but also decreases the number of necessary optical components, so minimizing the risk of errors introduced by them, for the study of a decoherent dynamics in optical setups. 

\section*{Acknowledgments}
F. L. acknowledge CAPES, B. M. and R. M. S. acknowledge CNPq; All the authors acknowledge support from Brazilian National Institute of Quantum Information (CNPq-INCT-IQ 465469/2014-0) and FAPESP 2021/14303-1.

\section*{Data Availability Statement}
The raw experimental results used in the tomography shown in the Figure~\ref{DPtomography} and~\ref{GADtomography} can be request to one of the correspondence e-mail address.
Results shown in Figure~\ref{DPlines} and~\ref{GADlines} used post-processing the results shown in Figure~\ref{DPtomography} and~\ref{GADtomography} to build the state and calculate Fidelity, Purity and Entanglement of the dynamic map, respectively.


\providecommand{\noopsort}[1]{}\providecommand{\singleletter}[1]{#1}%

\end{document}